\documentclass{article}
\usepackage{amsmath,epsfig,multirow,multicol,stfloats,graphicx,float, subfigure, mathrsfs,amsfonts}
\usepackage[preprint]{spconfa4}

\copyrightnotice{978-1-7281-1331-9/20/\$31.00 ©2020 IEEE}

\let\OLDthebibliography\thebibliography
\renewcommand\thebibliography[1]{
  \OLDthebibliography{#1}
  \setlength{\parskip}{0pt}
  \setlength{\itemsep}{0pt plus 0.3ex}
}

\begin{document}\sloppy

\def\x{{\mathbf x}}
\def\L{{\cal L}}

\title{STACKED CONVOLUTIONAL DEEP ENCODING NETWORK FOR VIDEO-TEXT RETRIEVAL}
%

\name{\renewcommand{\thefootnote}{\fnsymbol{footnote}}
Rui Zhao, Kecheng Zheng, Zheng-jun Zha\footnote[1]{Corresponding author.}}

\address{\begin{tabular}{c}University of Science and Technology of China\\
\{rzhao62, zkcys001\}@mail.ustc.edu.cn; zhazj@ustc.edu.cn\end{tabular}}

\maketitle

\renewcommand{\thefootnote}{\fnsymbol{footnote}}
\footnotetext[1]{Corresponding author.}
\begin{abstract}
Existing dominant approaches for cross-modal video-text retrieval task are to learn a joint embedding space to measure the cross-modal similarity. However, these methods rarely explore long-range dependency inside video frames or textual words leading to insufficient textual and visual details. In this paper, we propose a stacked convolutional deep encoding network for video-text retrieval task, which considers to simultaneously encode long-range and short-range dependency in the videos and texts. Specifically, a multi-scale dilated convolutional (MSDC) block within our approach is able to encode short-range temporal cues between video frames or text words by adopting different scales of kernel size and dilation size of convolutional layer. A stacked structure is designed to expand the receptive fields by repeatedly adopting the MSDC block, which further captures the long-range relations between these cues. Moreover, to obtain more robust textual representations, we fully utilize the powerful language model named Transformer in two stages: pretraining phrase and fine-tuning phrase. Extensive experiments on two different benchmark datasets (MSR-VTT, MSVD) show that our proposed method outperforms other state-of-the-art approaches.
\end{abstract}
\section{Introduction}
\label{sec:intro}

Cross-modal retrieval is a challenging area of research in the vision and language community. Giving one instance from either modality, it aims at identifying and retrieving corresponding correct instance from the other modality. While image-text retrieval has achieved much progress in recent years, video-text retrieval still remains further exploration. Compared with image, video contains richer and more complex information, leading to be difficult to represent these complex information due to the noise and irrelevant background. Meanwhile, just as Figure \ref{Fig,dataset} shows, for the video-text retrieval, a video is composed of several small events in which the corresponding sentence focuses on different parts of these events. Thus, both short-range and long-range dependency are necessary to be considered, which provide various semantic cues for the matching calculation between video and text. Besides, while using the textual description to describe the same event in a video, the words in a sentence may be variable. Hence, it is also significant to explore the way to efficiently capture the reliance among words.

\begin{figure}[t]
  \centering
  \includegraphics[width=8cm,height=7cm]{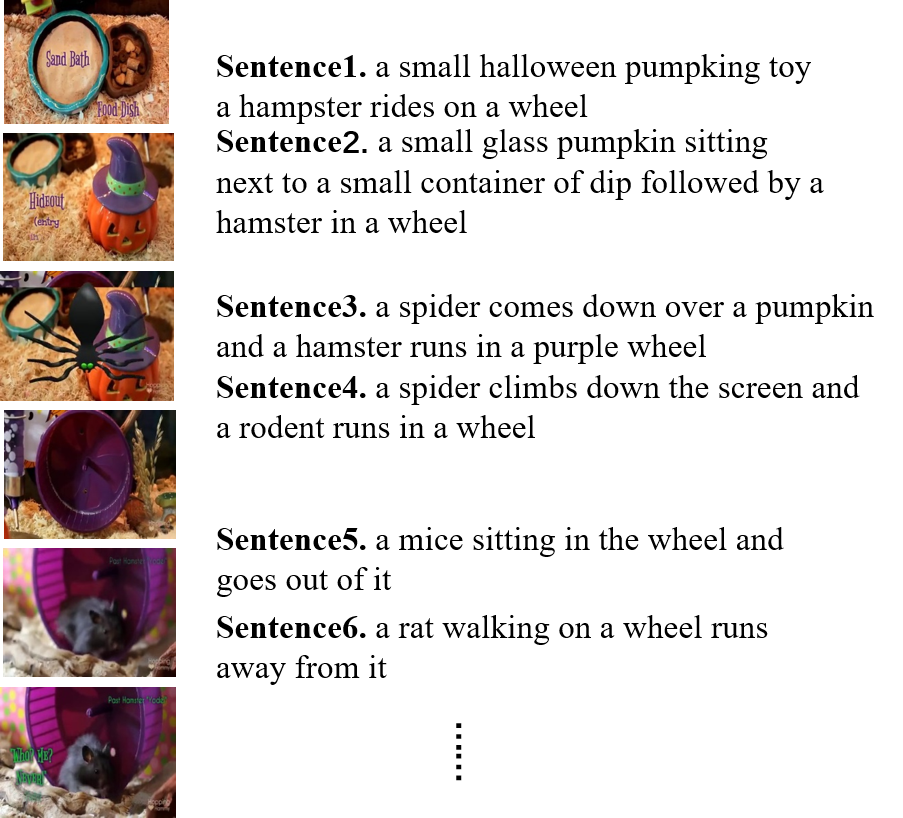}
  \caption{Illustration of a instance in MSR-VTT dataset, which includes some video frames and the corresponding textual descriptions. We can observe that some textual sentences(i.e. Sentence1 to Sentence4) describe the whole events in the video, which represents the long-range dependency. Meanwhile, Sentence5 and Sentence6 focus on the short-term event which is relevant to local frames in the video. It is worth mentioning that the word representations can be variable when the sentence describes the same events among video.
  }
  \label{Fig,dataset}
\end{figure}

To address the above problems, our work mainly focuses on the temporal cues modeling for both video and text. Convolutional neural network (CNN) is capable of aggregating spatial context and producing discriminative representation, thus is widely used in many tasks, such as action recognition, visual segmentation and so on. Especially, dilated spatial pyramid convolution \cite{yu2015multi} is effectively used in image segmentation for the spatial context modeling. Inspired by this, we extend it to exploit temporal semantic cues and relation learning for both video and text to produce deep discriminative features for retrieval.

Second, to better learn the word embedding, we fully utilize the superiority of Transformer for language modeling, which conducts position encoding and multi-head attention mechanism. In the first stage, we train the sentences by BERT in a unsupervised way, which randomly masks out some of the words in the sentence. The objective of BERT is to predict the masked words based only on the context, and thus the contextual information between words are captured inside the word embedding, which is then fine-tuned by the Transformer encoder in the second stage by the retrieval loss.

In this paper, a stacked convolutional deep encoding network is proposed to boost the video-text retrieval performance. The main contributions of this paper are summarized as follows: (1) We generalize the dilated spatial pyramid convolution to multi-scale dilated temporal convolution (MSDC) for capturing short-range temporal cues at the multiple scales. A stacked structure is designed to expand receptive fields by repeatedly adopting the MSDC block, which further captures the long-range relations between these cues. The Stacked MSDC can finally produce deep discriminative semantic cues for both video and text. (2) Unlike previous works that use word2vec to embed word and then encoded by RNN, we explore to pretrain word embedding with BERT and fine-tune with Transformer, which are powerful for language modeling. (3) Extensive experiments on two benchmark datasets show the effectiveness of our proposed method compared with the state-of-the-art methods.

\section{RELATED WORK}

The problem of vision and text retrieval can be divided into two categories: image-text retrieval and video-text retrieval. Previous methods for image-text retrieval mainly focus on a basic procedure:(1) extracting features from static image via deep convolutional neural network, (2) embedding words through word2vec and then encoded by RNN, and (3) measuring their similarity in a joint embedding space with a ranking loss to determine whether the input image and text are matched. Recently, there are also other interesting ideas to improve the performance, such as  cross-attention based models. The intuition is that different image-text pair may attend to each other in different local parts. 
\cite{wang2019camp} proposed to make attending between words and image regions symmetrically and exploit message-passing between two modalities. While cross-attention based methods are shown to be effective, but the computational overhead is $o(n^2)$ because each query instance should be encoded with all the reference instance, which may be time-consuming in practical application especially for large-scale datasets.
  
Similar to image-text retrieval, most video-text retrieval also conduct joint space learning. The basic procedure of video-text retrieval is also similar except that video feature is captured frame by frame via deep convolutional neural network (CNN) and then aggregated by RNN. \cite{mithun2018learning} also exploits to additionally use multi-modal video features for information complement, such as activity feature and audio feature, which are then fused into a single space or an extra space. \cite{dong2019dual} is the most related work to us, which proposed dual-encoding network with three level features for video-text retrieval. Our work focuses on the deep discriminative feature learning with temporal cues encoding and text embedding modeling and thus do not consider additional video features or require instance interaction.
Recent work \cite{zha2020adversarial} and \cite{zha2019context} also conduct discriminative feature learning for cross-modal tasks.

\section{OUR APPROACH}
In this section, we firstly present the overall architecture of the proposed approach, and then introduce each component of this architecture in the following subsection. The framework of the proposed stacked convolutional deep encoding network is shown in Figure \ref{Fig,model}. 

\subsection{Video-side Encoder}
Given a video, there are a large number of similar frames that contain redundant information. So we uniformly sample a sequence of $N$ frames with a interval of 0.5 second to represent the salient content of a video as follow \cite{mithun2018learning} and \cite{dong2019dual}. And then a ResNet-152 \cite{he2016deep} pretrained on ImageNet is adopted to extract the feature that represents each frame. In the end, each video is represented as $\boldsymbol{V}=\{\boldsymbol{v_1},\boldsymbol{v_2},\cdots,\boldsymbol{v_N}\}$ with $\boldsymbol{v_i}\in\mathbb{R}^{2048}$.

To capture the temporal information between video frames, we then feed $\boldsymbol{V}$ into a single-layer bidirectional GRU (bi-GRU \cite{schuster1997bidirectional}) with 512-dimensional hidden states to process the whole video. For each frame, we concatenate the hidden states from the two directional GRU as the global temporal representation:
\begin{equation}
\boldsymbol{\overrightarrow{h_i}} = \overrightarrow{GRU}(\boldsymbol{\overrightarrow{h_{i-1}}},\ \boldsymbol{v_i}),\quad\boldsymbol{\overleftarrow{h_i}} = \overleftarrow{GRU}(\boldsymbol{v_i},\ \boldsymbol{\overleftarrow{h_{i-1}}}),
\end{equation}
\begin{equation}
\boldsymbol{f_{v_i}}=concate(\boldsymbol{\overrightarrow{h_i}},\ \boldsymbol{\overleftarrow{h_i}}),\ i\in\left\{1,2,\cdots,N\right\},
\end{equation}
where $\boldsymbol{h_i}$ denotes to the $i$-th hidden state of bi-GRU.

Finally, we obtain a sequence of feature map $\boldsymbol{F_{vg}}=\left\{\boldsymbol{f_{v_1}},\boldsymbol{f_{v_2}},\cdots,\boldsymbol{f_{v_N}}\right\}\in \mathbb{R}^{N\times1024}$ from the bi-GRU, which refers to the representation of the each frame in the video. We then apply mean pooling operation on $\boldsymbol{F_{vg}}$:
\begin{equation}
\boldsymbol{f_{vg}} = \frac{1}{N}\sum\limits_{i=1}^N \boldsymbol{f_{v_i}},
\end{equation}
where $\boldsymbol{f_{vg}}$ denotes the global representation of the video.

\begin{figure*}[htbp]
\small
  \centering
  \includegraphics[height=6.8cm,width=16.5cm]{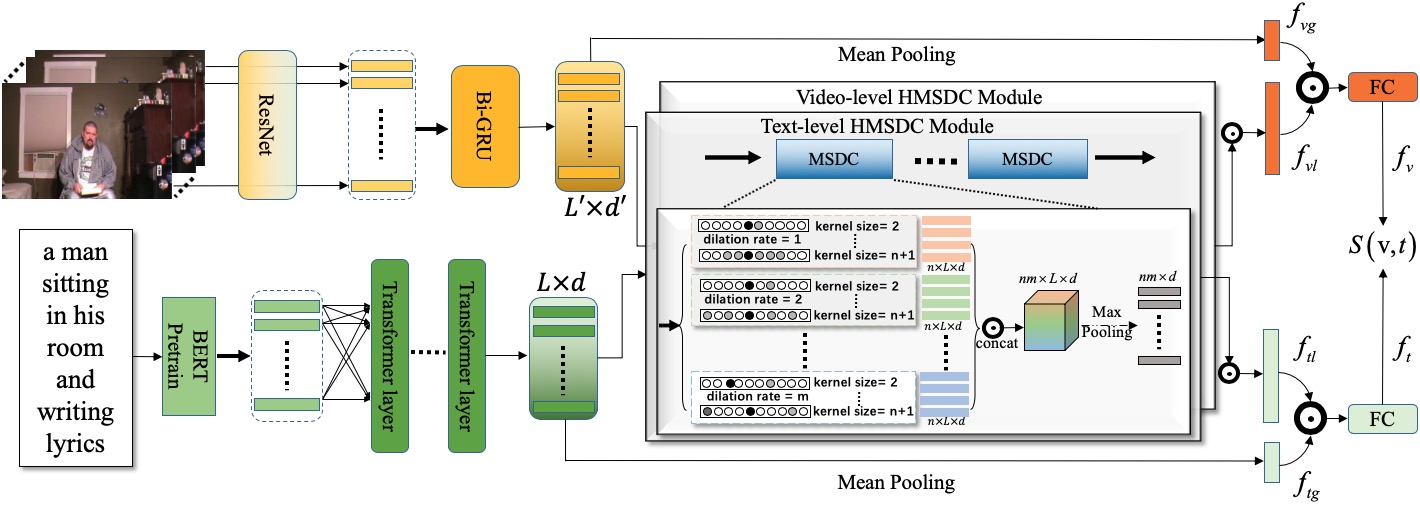}
  \caption{Illustration of the proposed framework. The network takes a pair of video and text sequences as inputs. The visual and textual features of the inputs are first extracted by their corresponding backbone. Then the global feature of video is encoded by bi-GRU while text is encoded by Transformer layer. Finally, the proposed Stacked Multi-scale Dilated Convolution module is apply on them to extract local temporal cues. The global and local features are concatenated to represent the inputs. They are finally projected into a joint space to learn the similarity between them.}
  \label{Fig,model}
\end{figure*}

\subsection{Text-side Encoder}
For text embedding, previous works~\cite{mithun2018learning,dong2019dual} usually embed a word by a traditional word2vec~\cite{mikolov2013efficient} model and then construct the contextual representation relaying on a single RNN. In this way, the textual representations are not discriminative enough. Recently, a great progress has been made in the field of natural language processing (NLP): a language representation model BERT \cite{devlin2018bert} is proposed and has proven to be effective in many NLP tasks. Motivated by this, our textual data is pretrained by BERT. To be specific, the BERT is trained to predict masked words inside a sentence based only on the context in an unsupervised way. In this way, we can capture the discriminative contextual information between words. After pretraining, the word embedding is then fine-tuned by Transformer module~\cite{vaswani2017attention} in retrieval task. It is more effective and efficient than RNN, because of its superiority in language modeling.

Given a sentence with length $M$, we can obtain the $M$ fixed embeddings from the pretrained BERT. Specifically, we average the outputs of the last four layer from BERT as the embeddings. We enable to achieve the $M$ fixed embeddings $\boldsymbol{T}=\{\boldsymbol{t_1},\boldsymbol{t_2},\cdots,\boldsymbol{t_M}\}$ with $\boldsymbol{t_i}\in \mathbb{R}^{768}$. After obtaining the embeddings of a sentence, we send them into Transformer module to extract the global encoding of words, which employ multi-head self-attention to capture the abundant contextual information. We formulate this as:
\begin{equation}
\begin{aligned}
  {\boldsymbol{f_{t_i}}}^k=Transformer^{(k)}({\boldsymbol{f_{t_i}}^{k-1}}),\\
  i\in\{1,2,\cdots,M\},\ k\in\{1,2,\cdots,K\},
\end{aligned}
\end{equation}
where $K$ is layer number of Transformer, $M$ refers to the length of a sentence and ${\boldsymbol{f_{t_i}}}^0=\boldsymbol{t_i}$.

The global temporal feature map of a sentence is represented as $\boldsymbol{F_{tg}}=\{\boldsymbol{f_{t_1}},\boldsymbol{f_{t_2}},\cdots\,\boldsymbol{f_{t_M}}\}\in \mathbb{R}^{M\times768}$. Same to video side, global representation of a sentence is produced by:
\begin{equation}
  \boldsymbol{f_{tg}}=\frac{1}{M}\sum\limits_{i=1}^M \boldsymbol{f_{t_i}},
\end{equation}

\subsection{Stacked Multi-scale Dilated Convolution}

In addition to global information, local cues are proved to be also necessary in the field of computer vision and natural language processing. So we need to design a module to simultaneously capture the global and local cues. In this paper, we propose a Stacked Multi-scale Dilated Convolutional (SMSDC) module which can be inserted into video or text processing model. This module consists of two steps: the first step is designed to capture the local temporal cues among consecutive frames and skipped frames, while the second step is designed to capture the complex long-range relations among these local temporal cues. Our intuition is that diverse and refined information of an instance can be derived in this way.

Spatial convolution is capable of modeling the spatial context within an image, which is widely used in many tasks such as Image Segmentation. Inspired by this, we implement a multi-scale dilated convolution (MSDC) which represents local temporal feature learning and relation learning both in videos and texts.

The MSDC takes global feature map $\boldsymbol{F_g}$ (either of $\boldsymbol{F_{vg}}$ and $\boldsymbol{F_{tg}}$) as input and outputs a local feature vector $\boldsymbol{f_l}$ ($\boldsymbol{f_{vl}}$ for $\boldsymbol{F_{vg}}$ and $\boldsymbol{f_{tl}}$ for $\boldsymbol{F_{tg}}$). After obtaining this global feature map $\boldsymbol{F_g}\in \mathbb{R}^{L\times d}$, we adopt the dilated convolution (DC) operation which is formulated as:
\begin{equation}
\begin{aligned}
\boldsymbol{f^{(r,w)}(t)}=\sum_{i=1}^w \boldsymbol{F_g[t+r\cdot i]}\times \boldsymbol{W_{[i]}^{(r,w)}},\ t=1,2,\cdots,L
\end{aligned}
\end{equation}
where $\boldsymbol{W_{[i]}^{(r,w)}}$ denotes the parameters in the the dilated convolution, $w$ is the kernel size and $r$ is the dilated size.

$\boldsymbol{F_{DC}^{(r,w)}}=\{\boldsymbol{f^{(r,w)}(t)}\} \in \mathbb{R}^{L\times d}$ is the collection of output features of DC. The dilated convolution adaptively fuses the local frames or words together for constituting the different local events. These events contain the different local temporal cues.


To capture more different semantic cues, we extend DC to a multi-scale dilated (MSDC) form with different kernel scale and dilation size, which can capture local temporal cues in terms of different receptive field. Defining $w_n\in \{2,\cdots,n+1\}$ be the kernel scale set and $r_m\in \{1,\cdots,m\}$ be the dilation size set. $\boldsymbol{F_{MSDC}}=concate(\boldsymbol{F_{DC}^{(r_1,w_1)}};\boldsymbol{F_{DC}^{(r_1,w_2)}};\cdots;\boldsymbol{F_{DC}^{(r_m,w_n)}}) \in \mathbb{R}^{nm\times L \times d}$ is the output of MSDC. To represent each input to a vector, we apply nonlinear activation $\sigma$ and maxpooling along the time dimension:

\begin{equation}
  \boldsymbol{F_{MSDC}}=maxpooling(\sigma(\boldsymbol{F_{MSDC}})) \in \mathbb{R}^{nm\times d},
\end{equation}

We then stack another MSDC on $\boldsymbol{F_{MSDC}}$ to model the long-range relations among the $nm$ different local semantic features, producing a discriminative feature representation $\boldsymbol{F_{SMSDC}} \in \mathbb{R}^{nm\times d}$.

To fuse the $nm$ features into a single vector, we simply apply concatenation fusion to get the final local representation:
\begin{equation}
\begin{aligned}
  \boldsymbol{f_l} = concate(\boldsymbol{F_{SMSDC}[0]},\cdots,\boldsymbol{F_{SMSDC}[nm-1]}),
\end{aligned}
\end{equation}
where $\boldsymbol{f_l}\in \mathbb{R}^{nmd}$ refers to the output of our proposed SMSDC, which is used to calculate the similarity between video and text.

\subsection{Model Learning}
After encoding the video and text, we can obtain the global and local features of them, respectively. Then, we apply a concatenation operation to fuse them in order to provide more semantic cues for matching calculation between video and text.
\begin{equation}
\boldsymbol{f_{v_{gl}}}=concate(\boldsymbol{f_{vg}}, \boldsymbol{f_{vl}}),\quad \boldsymbol{f_{t_{gl}}}=concate(\boldsymbol{f_{tg}}, \boldsymbol{f_{tl}}),
\end{equation}

In order to calculate the similarity between video and text, we embed them into a joint space, in which the embedding of the positive video-text pairs should be close to each other while the negative pairs should be far away.

Following \cite{dong2019dual}, we use a Fully Connected (FC) layer following with a Batch Normalization(BN) layer as the final embedding layer, which can be formulated as:
\begin{equation}
\begin{aligned}
\boldsymbol{f_v}=BatchNorm(\boldsymbol{W_v}\boldsymbol{f_{v_{gl}}}+b_v),\\
\boldsymbol{f_t}=BatchNorm(\boldsymbol{W_t}\boldsymbol{f_{t_{gl}}}+b_t),
\end{aligned}
\end{equation}
where $\boldsymbol{W_v}$, $\boldsymbol{W_t}$, $b_v$, $b_t$ are weights and biases.

With the final embedding of video and text as described above, the similarity of $f_v$ and $f_t$, denoted as $S(v,t)$, can then be computed as the cosine of the angle between them. The similarity should be subject to the model parameters during the optimization. We utilize bidirectional max-margin hard ranking loss to optimize the model, which is the state-of-the art loss and widely used in vision and language matching task. The loss of this optimization problem can then be written as:
\begin{equation}
\begin{aligned}
\mathcal{L}=max(0,\alpha-S(v,t)+S(v,t^-))\\
+max(0,\alpha-S(v,t)+S(v^-,t)),
\end{aligned}
\end{equation}
where $\alpha$ is the margin constant and is set as a hyperparameter. During the training, hard negative sample is utilized within a batch to penalize the model: $t^-$ is the negative text for $v$ while $v^-$ is the negative video for $t$.

The whole model is trained end-to-end to minimize $\mathcal{L}$ except that the image feature and word embedding are respectively extracted by pretrained CNN and pretrained BERT, which are all fixed.

\section{EXPERIMENTS}
We first introduce our experiment setup and implementation details. Then, we present the experimental result and the comparison with previous work. After that, ablation study is shown.

\subsection{Experimental Settings}

\textbf{Datasets.} In this paper, we conduct experiments on two benchmark datasets: MSR-VTT \cite{xu2016msr} and MSVD \cite{chen2011collecting} to evaluate the performance of our proposed framework. MSR-VTT and MSVD are two challenging datasets in the filed of video question answering, video captioning and video-text retrieval. MSR-VTT consists of 10000 video clips , each of which is annotated with 20-sentence descriptions. 
MSVD is a small dataset, which contains 1970 videos, while each of them has around 40 sentences. Note that, there are two kinds of sentence constructing strategies in the previous work:  JMET \cite{pan2016jointly} uses all the ground-truth sentences while LJRV \cite{otani2016learning} randomly samples 5 ground-truth sentences for each video. We follow the latter strategy. The splits of train, validation and test for the two datasets are consistent to previous work.

\begin{table*}[ht]
\small
  \begin{center}
    \caption{Performance comparison with other state-of-the-art methods on MSR-VTT dataset. Higher R@K, mAP and lower MedR is better. Sum of Recalls (RSum) indicates the overall performance. Our method achieves the best. }\label{tab:msrvtt}
    \begin{tabular}{c|ccccc|ccccc|c}
      \hline
      \multirow{2}{*}{Method} & \multicolumn{5}{c|}{Text-to-Video Retrieval} & \multicolumn{5}{c|}{Video-to-Text Retrieval} & \multirow{2}{*}{RSum} \\ \cline{2-11}
      & R@1    & R@5     & R@10   & MedR   & mAP     & R@1     & R@5    & R@10   & MedR   & mAP     &                                 \\ \hline
      W2VV\cite{dong2016word2visualvec}                    & 1.8    & 7.0     & 10.9   & 193    & 0.052   & 9.2     & 25.4   & 25.4   & 24     & 0.050   & 90.3                            \\ \hline
      VSE\cite{mithun2018learning}                    & 5.0    & 16.4    & 24.6   & 47     & -       & 7.7     & 20.3   & 31.2   & 28     & -       & 105.2                           \\ \hline
      VSE++\cite{mithun2018learning}                    & 5.7    & 17.1    & 24.8   & 65     & -       & 10.2    & 25.4   & 35.1   & 25     & \_      & 118.3                           \\ \hline
      Mithun et al.\cite{mithun2018learning}            & 5.8    & 17.6    & 25.2   & 61     & -       & 10.5    & 26.7   & 35.9   & 25     & -       & 121.7                           \\ \hline
      W2VViml\cite{dong2019dual}                 & 6.1    & 18.7    & 27.5   & 45     & 0.131   & 11.8    & 28.9   & 39.1   & 21     & 0.058   & 132.1                           \\ \hline
      DualEncoding\cite{dong2019dual}            & 7.7    & 22.0    & 31.8   & 32     & 0.155   & 13.0    & 30.8   & 43.3   & 15     & 0.065   & 148.6                           \\ \hline
      \textbf{Ours}                    & \textbf{8.8}    & \textbf{25.5}    & \textbf{36.5}   & \textbf{22}     & \textbf{0.174}   & \textbf{14.0}    & \textbf{33.1}   & \textbf{44.9}   & \textbf{14}     & \textbf{0.076}   & \textbf{162.8}                           \\ \hline
    \end{tabular}
  \end{center}
\end{table*}
\begin{table*}[ht]
\small
  \begin{center}
    \caption{Performance comparison with other state-of-the-art methods on MSVD dataset, using the same partition in LJRV.}\label{tab:msvd}
    \begin{tabular}{c|ccccc|ccccc|c}
      \hline
      \multirow{2}{*}{Method} & \multicolumn{5}{c|}{Text-to-Video Retrieval} & \multicolumn{5}{c|}{Video-to-Text Retrieval} & \multirow{2}{*}{RSum} \\ \cline{2-11}
      & R@1     & R@5    & R@10   & MedR   & MeanR   & R@1     & R@5    & R@10   & MedR   & MeanR   &                                 \\ \hline
      ST\cite{kiros2015skip}                     & 2.6     & 11.6   & 19.3   & 51     & 106     & 3.0     & 10.9   & 17.5   & 77.0   & 241.0   & 64.9                            \\ \hline
      LJRV\cite{otani2016learning}                    & 7.7     & 23.4   & 35     & 21     & 49.1    & 9.9     & 27.1   & 38.4   & 19.0   & 75.2    & 141.5                           \\ \hline
      W2VV-ResNet\cite{mithun2018learning}             & -       & -      & -      & -      & -       & 17.9    & 39.6   & 51.3   & 11.0   & 57.6    & -                               \\ \hline
      Mithun et al.\cite{mithun2018learning}           & 15.0    & 40.2   & 51.9   & 9.0    & 45.3    & 20.9    & 43.7   & 54.9   & 7.0    & 56.1    & 226.7                           \\ \hline
      \textbf{Ours}                    & \textbf{18.2}    & \textbf{44.0}   & \textbf{57.8}   & \textbf{7.0}    & \textbf{38.6}    & \textbf{23.2}    & \textbf{48.2}   & \textbf{62.5}   & \textbf{6.0}    & \textbf{38.8}    & \textbf{253.9}                           \\ \hline
    \end{tabular}
  \end{center}
\end{table*}

\begin{table*}[ht]
\small
\begin{center}
\caption{Ablation study on MSR-VTT dataset to investigate contributions of different components of our model}\label{tab:msrvtt_ab}
\begin{tabular}{c|ccccc|ccccc|c}
\hline
\multirow{2}{*}{Method} & \multicolumn{5}{c|}{Text-to-Video Retrieval} & \multicolumn{5}{c|}{Video-to-Text Retrieval} & \multirow{2}{*}{RSum} \\ \cline{2-11}
                        & R@1    & R@5     & R@10   & MedR   & mAP     & R@1     & R@5    & R@10   & MedR   & mAP     &                       \\ \hline
baseline                & 6.2    & 19.6    & 29.2   & 36     & 0.136   & 9.8     & 27.1   & 38.2   & 19     & 0.056   & 130.2                 \\ \hline
+BERT                   & 7.0    & 21.6    & 31.8   & 29     & 0.151   & 11.6    & 30.3   & 41.0   & 17     & 0.064   & 143.3                 \\ \hline
+Transformer                    & 7.4    & 22.5    & 33.0   & 27     & 0.156   & 12.3    & 30.5   & 41.2   & 17     & 0.069   & 146.9                 \\ \hline
+MSDC                   & 8.5    & 24.9    & 35.6   & 23     & 0.173   & 13.6    & 32.4   & 43.8   & 14     & 0.075   & 158.7                 \\ \hline
+SMSDC                  & 8.8    & 25.5    & 36.5   & 22     & 0.176   & 14.0    & 33.1   & 44.9   & 14     & 0.076   & 162.8                 \\ \hline
\end{tabular}
\end{center}
\end{table*}

\textbf{Evaluation Metric.} Following prior work on vision-text matching task, we use the standard rank-based criteria: R@K (Recall at rank K, K=1, 5, 10), MedR (median Rank), MeanR (mean Rank) and mAP (mean Average Precision), to report our retireval performance. $RSum=R@1+R@5+R@10$ is calculated to compare the overall preformance. Note that, due to the missing result in prior work, we only report mAP for MSR-VTT and MeanR for MSVD.

\textbf{Implementation Details.} We set the margin $\alpha=0.2$ in the rank loss and the dimension of the embedding space in equation $(10)$ is set to 2048. (n, m) pair in SMSDC is empirically set to (4, 2) for video and (3, 2) for text. The Transformer layer number K is set to 3. We keep the batch size to 64. Learning rate is initialized to $5\times10^{-5}$ and is decreased by a factor of 2 once the performance does not increase in three consecutive epochs. SGD with Adam is adopted as the optimizer. The maximal number of training epoch is set to 30. The best model is chosen by sum of recalls.

\subsection{Compared with state-of-the-art}

Table \ref{tab:msrvtt} and Table \ref{tab:msvd} show the experimental results and comparisons with previous methods on the two datasets respectively. All the results are cited from its original papers if available. We can see that our proposed method performs best and consistently outperforms state-of-the-art methods in both text-to-video and video-to-text retrieval. It verifies the effectiveness of our proposed method. The performance of video-to-text is higher than text-to-video because one video is paired with several sentence and we take the most similar one as the rank. Sum of recalls is increased from 148.6 to 162.7 on MSR-VTT dataset and 226.7 to 253.9 on MSVD dataset. Note that, we also use the source code and original settings of \cite{dong2019dual} to test on MSVD, the sum of recalls is around 241, which is still much lower than ours. The relative improvement is smaller on MSVD dataset compared with a same method, the reason may be that the length of videos are longer and the number of sentence is more and thus can provide more semantic information. 

\subsection{Ablation Study}
We also conduct experiments on MSR-VTT dataset to evaluate the effectiveness of different components in the proposed method, including the text side embedding modeling and the proposed convolution module. The experimental results are shown in table \ref{tab:msrvtt_ab}, in which the baseline is both video feature and text feature encoding by Bi-GRU with word2vec embedding. Then the word embedding and Bi-GRU of text respectively replaced by BERT pretrained embedding and Transformer. Compared with baseline method, BERT pretraining with Transformer fine-tuning is effective. Then we add MSDC on both video and text, we can see the result is improved, which demonstrates the effectiveness of capturing dense temporal cues. SMSDC further improves the performance, which further captures the relation between temporal cues and result in a refined and discriminative feature. 
\section{CONCLUSION}

In this paper, we have proposed a stacked convolutional deep encoding network to capture more discriminative semantic cues for video-text retrieval. A SMSDC module within our method is able to encode short-range temporal cues and long-range relations between them by adopting convolutions with different kernel size and dilation size. In addition, a more robust textual representation is obtained by fully utilizing the superiority of Transformer in language modeling. Through extensive experiments, we demonstrate that our method outperforms state-of-the-art video-text retrieval methods.

\section{ACKNOWLEDGMENTS}
This work was supported by the National Key R\&D Program of China under Grant 2017YFB1300201, the National Natural Science Foundation of China (NSFC) under Grants 61622211, U19B2038 and 61620106009 as well as the Fundamental Research Funds for the Central Universities under Grant WK2100100030.

{
\small
\bibliographystyle{IEEEbib}
\bibliography{icme2020template}
}

\end{document}